\documentclass[11pt]{article}
\usepackage{epsfig,amscd,amssymb}
\begin{document}

\def\wgta#1#2#3#4{\hbox{\rlap{\lower.35cm\hbox{$#1$}}
\hskip.2cm\rlap{\raise.25cm\hbox{$#2$}}
\rlap{\vrule width1.3cm height.4pt}
\hskip.55cm\rlap{\lower.6cm\hbox{\vrule width.4pt height1.2cm}}
\hskip.15cm
\rlap{\raise.25cm\hbox{$#3$}}\hskip.25cm\lower.35cm\hbox{$#4$}\hskip.6cm}}
\def\wgtb#1#2#3#4{\hbox{\rlap{\raise.25cm\hbox{$#2$}}
\hskip.2cm\rlap{\lower.35cm\hbox{$#1$}}
\rlap{\vrule width1.3cm height.4pt}
\hskip.55cm\rlap{\lower.6cm\hbox{\vrule width.4pt height1.2cm}}
\hskip.15cm
\rlap{\lower.35cm\hbox{$#4$}}\hskip.25cm\raise.25cm\hbox{$#3$}\hskip.6cm}}

%
%
%
%
\title{BPS kinks in the Gross-Neveu model}

\author{Paul Fendley$^1$ and Hubert Saleur$^2$\\
\smallskip\\
$^1$ Department of Physics\\
University of Virginia\\
Charlottesville, VA 22904-4714\\
{\tt fendley@virginia.edu}\\
\smallskip\\
$^2$ Department of Physics and Astronomy\\
University of Southern California\\
Los Angeles, CA 90089}

\maketitle

\begin{abstract}

We find the exact spectrum and degeneracies for the Gross-Neveu model
in two dimensions.  This model describes $N$ interacting Majorana
fermions; it is asymptotically free, and has dynamical mass generation
and spontaneous chiral symmetry breaking. We show here that the
spectrum contains $2^{N/2}$ kinks for any $N$.  The unusual $\sqrt{2}$
in the number of kinks for odd $N$ comes from restrictions on the
allowed multi-kink states. These kinks are the BPS states for a
generalized supersymmetry where the conserved current is of dimension
$N/2$; the $N=3$ case is the ${\cal N}=1$ supersymmetric sine-Gordon
model, for which the spectrum consists of $2\sqrt{2}$ kinks.  We find
the exact $S$ matrix for these kinks, and the exact free energy for
the model.

\end{abstract}

\newpage

\section{Introduction}

The Gross-Neveu model describes interacting fermions in two
dimensions.  It has no gauge fields or gauge symmetry, yet it exhibits
much of the behavior of gauge theories in four dimensions.  The
coupling constant is naively dimensionless, but radiative corrections
result in a mass scale to the theory. Thus the theory is
asymptotically free and strongly interacting in the infrared. A
(discrete) chiral symmetry is spontaneously broken, which gives the
fermions mass. In condensed-matter language, the interaction is
marginally relevant, and there is no non-trivial fixed point.

The Gross-Neveu model consists of $N$ Majorana fermions
$\psi^i$, $\bar{\psi}^i$, with $i=1\dots N$. 
The action is
\begin{equation}
S= \int d^2 z \left[\psi^i \bar\partial \psi^i + \bar\psi^i \partial
\bar\psi^i + g (\bar\psi^i \psi^i)(\bar\psi^j \psi^j)\right]
\label{action}
\end{equation}
where repeated indices are summed over. At the critical point $g=0$,
the fermions are free, with $\psi^i$ a function only of $z$ and $\bar
\psi^i$ a function only of $\bar z$.  At this critical point in two
dimensions, the fermions have left and right dimensions $(1/2,0)$ for
$\psi$ and $(0,1/2)$ for $\bar\psi$, so the coupling $g$ is naively
dimensionless. However, the beta function for this interaction is
non-vanishing \cite{GN}. For $g>0$, the trivial
free-fermion fixed point is unstable, and a mass scale is
generated. We denote this mass scale $M$.  In this paper we study
$N\ge 3$; for $N=2$ the model reduces to the well-known massless
Thirring model (the Luttinger model in the condensed-matter
literature), and for $N=1$ it is free.  The action (\ref{action}) is
invariant under the symmetry group $O(N)$. In a separate paper, we will
discuss generalizations to other symmetries.

Despite a huge number of papers discussing various aspects of this
model, there remained an important unanswered question: what is the
spectrum of particles? Just because the fields are known does not mean
the spectrum is known: there may be kink states or bound states, and
in some cases there are no one-particle states corresponding to the
fields themselves.  In particular, it was long known that for even $N$
that the spectrum contained $2^{N/2}$ kink states in the two spinor
representations of the $SO(N)$ symmetry algebra \cite{Witten}. The
arguments which lead to the existence of kinks for even $N$ apply
equally well to odd $N$, but it was not clear how many such kinks
there were when $N$ is odd. It was shown in \cite{OGW} that there is
no consistent $S$ matrix for a single spinor representation, and by
utilizing a generalized supersymmetry, Witten showed that the kinks
are in a reducible representation of $SO(2P+1)$. He went on to
conjecture that they are in two copies of the $2^P$-dimensional spinor
representation.

In this paper we will answer this question and at long last complete
the computation of the mass spectrum. We show that
there are $\sqrt{2}$ copies of the spinor representation when $N$ is odd:
the number of kinks for any $N$ is $2^{N/2}$. A non-integer number of kinks
means that there are restrictions on the allowed multi-kink states; we
give a precise definition below. These kinks are in BPS representations
of the generalized supersymmetry. For $N=3$, the model is equivalent
to the ${\cal N}=1$ supersymmetric sine-Gordon model, so our result
shows that these unusual BPS kinks appear even for ordinary supersymmetry.
To confirm our claims we compute the exact $S$ matrix and free energy
of this model, and find agreement with the known results
in the ultraviolet limit.

\section{The symmetries}

To determine the spectrum and $S$ matrices of the Gross-Neveu model,
we need to understand the symmetry structure of the model in depth.
There are actually four different symmetries which we utilize in this
paper. These are all discussed in \cite{Witten}; this section is
basically a review of these results.

\bigskip{\bf $O(N)$ symmetry}

\smallskip
\noindent The action has a global
$O(N)$ symmetry $\psi^i \to U^{ij} \psi^j$ and $\bar\psi^i \to
U^{ij\, T} \bar\psi^j$. The matrix $U$ must be an element of $O(N)$
because the Majorana fermions are real. The existence of this global
symmetry means that the particles of the model must transform in
representations of this symmetry.

\bigskip{\bf Spontaneously-broken chiral symmetry}

\smallskip\noindent
The action (\ref{action}) has a ${\bf Z}_2$ chiral symmetry $\psi\to
-\psi$, $\bar\psi \to \bar\psi$.  However, for $g>0$, this symmetry is
spontaneously broken, because the fermion bilinear $\sigma\equiv
\bar\psi^i \psi^i$ gets an expectation value. This expectation value
results in a mass for the fermions. Note that this expectation value
does not break the $O(N)$ symmetry; a continuous symmetry
cannot be spontaneously broken in two spacetime dimensions.
Equivalently, the discrete parity symmetry $\psi(z,\bar z)\to
\bar\psi(\bar z,z)$ is spontaneously broken.

\bigskip{\bf Local conserved charges}

\smallskip\noindent When a model
possesses an infinite number of local conserved currents transforming
non-trivially under the Lorentz group, it is said to be
integrable. The integrability results in powerful constraints on the
$S$ matrix, which will be discussed below. It was shown in
\cite{Witten} that the action (\ref{action}) has at least one of these
conserved currents. When $g=0$, the energy-momentum tensor $T= \psi^i
\partial\psi^i$ obeys $\bar\partial T^n=0$ for any integer $n$. This
no longer holds for $g\ne 0$.  However, $\bar\partial T^2$ must have
dimension $5$ and Lorentz spin $3$. All operators
of this dimension and spin must be a total derivative:
$$\bar\partial T^2 = \bar\partial A + \partial B$$ Thus there is a
conserved current with components $(T^2-A,B)$ in the Gross-Neveu
model. This already requires that the scattering be factorizable, and
it seems very likely that there is a conserved current of dimension
$2n$ (and hence a charge of dimension $2n-1$) for all integer $n$.

\bigskip{\bf Generalized supersymmetry}

\smallskip\noindent
A very striking feature of the Gross-Neveu model shown in \cite{Witten}
is that it possesses an extra conserved current of spin $N/2$. 
This conserved current is
\begin{equation}
J= \epsilon_{i_1i_2\dots i_N} \psi^{i_1}\psi^{i_2}\dots \psi^{i_N} .
\label{current}
\end{equation}
Another conserved current $\bar J$ of spin $-N/2$ follows from
replacing $\psi$ with $\bar\psi$. To prove that this is a conserved
current even for $g\ne 0$, one shows that all possible
contributions to $\bar\partial J$ are themselves total derivatives.
In an equation,
$$\bar\partial J = \partial \widetilde J$$
For $N=2$, $\widetilde J$ vanishes
and this conserved current is the axial symmetry of the massless
Thirring model (Luttinger model). For $N=4$, the Gross-Neveu model
decouples into two sine-Gordon models, and the spin-2 current $J$ is
the difference of the energy-momentum tensors of the two theories. 

For odd $N=2P+1$, the results are much more surprising. For $N=3$, the
current is of dimension $3/2$, and generates supersymmetry
transformations. This seems odd in a theory of fermions, but by
bosonizing two of the three fermions, one indeed obtains the ${\cal
N}=1$ supersymmetric sine-Gordon model \cite{Witten}. For
general odd $N$, this results in a generalized supersymmetry. The full
current algebra of these currents seems quite tricky. It seems very
likely that at $g=0$ it is the $WB_P$ algebra studied in \cite{FL},
which indeed involves a current of spin $P+1/2$ along with those of
even integer spin.  This was explicitly worked out for the $P=2$ case
(when $g=0$) in \cite{Ahn}. This current algebra involves the spin-4
local current as well. These currents remain conserved
when $g\ne 0$, just like the supersymmetry currents do when $P=1$.

Luckily, to determine the $S$ matrix we do not need the full current
algebra: we need only to understand how the conserved charges act on
the particles. The conserved charge $Q$ of dimension $N/2 -1$ is
defined by $Q=\int dz J + \int d\bar z \widetilde J$, and likewise for
$\overline Q$. The operator $Q^2$ must be of dimension and Lorentz
spin $N-2$.  For example, for $N=3$, using the explicit form of the
charge gives $Q^2 \propto P_L$ and $\overline Q^2 = P_R$, with $P_L$
and $P_R$ the left and right momenta \cite{Aratyn}.  For general $N$,
this means that when $Q^2$ acts on a particle with energy $E$ and
momentum $p$,
$$Q^2 \propto (E+p)^{N-2},$$ since this is the only combination of
energy and momentum with the correct Lorentz properties.
It is convenient to define the rapidity $\theta$
so that a particle of mass $m$ has energy $E=m\cosh\theta$ and
momentum $p=m\sinh\theta$. 
The symmetry algebra acting on a particle for odd $N$ can then be written
$$
\{Q,Q\}= 2m^{N-2}e^{(N-2)\theta} \qquad\ \{\overline Q,\overline Q\} \propto
2m^{N-2} e^{-(N-2)\theta}$$
\begin{equation}\{Q,\overline Q\} = 2Z
\label{central}
\end{equation}
The central term $Z$ acts on the states as 
$Z|\alpha\rangle = m^{N-2}z_\alpha
|\alpha\rangle$. The (real) numbers $z_\alpha$ vanish
at $g=0$, but do not otherwise. As we will discuss, non-zero
$z_\alpha$ occur because
of the non-trivial boundary conditions on the kink states \cite{WO}.

\section{The kink spectrum of the Gross-Neveu model}

One interesting feature of the Gross-Neveu model is that the spectrum
is far more intricate than a glance at the action (\ref{action}) would
suggest. For any value of $N$, the spectrum includes kink states, as
pointed out in \cite{Zunpub,Witten}.  In fact, for $N=3$ and $N=4$,
the spectrum does not contain anything but the kink states.
The spectrum includes kink states because of the spontaneously broken
${\bf Z}_2$ chiral symmetry. The kink interpolates between the two
degenerate vacua; in two-dimensional spacetime there are domain walls
between regions of the two vacua. 
The main result of this paper is to answer the question: how many kink
states are there? 

The result requires first finding how the kink states transform under
the symmetries discussed in the last section.  We will find the
minimal number of particles required to transform under these
symmetries non-trivially. In section 4, we will work out the
$S$ matrix of these particles, and then in section 5
we will show that these
particles and $S$ matrices give the correct free energy.

Again, we follow the analysis of \cite{Witten} and study the kinks
semiclassically. The fermion bilinear $\sigma \equiv \bar\psi^i
\psi^i$ gets an expectation value $\sigma_0$; the kink is a
configuration of $\sigma$ interpolating between $+\sigma_0$ and
$-\sigma_0$. In the semi-classical approach, one treats $\sigma$ as a
classical background field, and quantizes the fermions in this
background. The fermions interact with the background by
the interaction 
$$g \sigma \bar\psi^i \psi^i$$ Thus in the semiclassical limit, this
problem is equivalent to quantizing $N$ Majorana fermions in a kink
background.  A famous result of Jackiw and Rebbi shows that the
Dirac equation with this
background possesses a single normalizable zero-energy
mode $f_0,\bar f_0$ \cite{JR}. 
The zero-energy solution is real, while the
finite-energy solutions are complex, so the zero-energy state is
created/annihilated by a real operator $b^i$, while the finite-energy
states are annihilated and created by $a_n^i$ and $a_n^{i\dagger}$
respectively. 
Thus the semiclassical expansion for $\psi^i$ and $\bar\psi^i$ is
\begin{eqnarray}
\nonumber
\psi^i &=& f_0 b^i + \sum_{n=0}^\infty 
\left[f_n a_n^i + f_n^* a_n^{i\dagger}\right]\\
\bar\psi^i &=& \bar{f}_0 b^i + \sum_{n=0}^\infty 
\left[\bar f_n a_n^i + \bar f_n^* a_n^{i\dagger}\right]
\label{semi}
\end{eqnarray}
The canonical commutation relations for $\psi^i$ require that
\begin{equation}
\{b^i,b^j\}=2\delta^{ij}
\label{clifford}
\end{equation}
This is called a Clifford algebra; 
the combinations $M^{ij} = [b^i,b^j]$ generate the $SO(N)$ algebra.
In this paper, when we use $O(N)$ we generally mean the symmetry group,
while $SO(N)$ refers to the resulting symmetry algebra.

The presence of these zero modes means that there must be more than
one kink state, because one can always act with any of the $b^i$ and
get a different state. In other words, denoting the kink states by 
$|\alpha\rangle$, some matrix elements
\begin{equation}
\langle \beta | b^i |\alpha\rangle
\label{matrixel}
\end{equation}
are non-vanishing. The kink states therefore form a representation
of the Clifford algebra.
The simplest possibility is that the $b^i$ 
act like the gamma matrices of $SO(N)$. A gamma matrix
$(\gamma^i)_{rs}$ obeys the Clifford algebra (\ref{clifford}),
and is an $SO(N)$ invariant where $i=1\dots N$ transforms
in the vector representation of $SO(N)$, while $r,s=1\dots 2^{int(N/2)}$
transform in the spinor representation(s). The kinks must
therefore be in the spinor representation(s) of $SO(N)$. 
This is effectively charge fractionalization, because the
vector representation appears in the tensor product of two
spinor representations. Thus the original fermions $\psi^i$
in the vector representation ``split'' into kinks in the
spinor representation(s).

The properties of the spinor representation(s) of $SO(N)$ are different
for even and odd $N$. For even $N$, there is an operator
$\gamma^{N+1}\equiv ib^1 b^2\dots b^N$ which commutes with all the
$SO(N)$ generators $M^{ij}$ but anticommutes with the Clifford algebra
generators.  Therefore the $2^{N/2}$-dimensional representation of the
Clifford algebra is not irreducible under the $SO(N)$ when $N$ is even.
It decomposes into irreducible representations with
$\gamma^{N+1}=\pm 1$. These representations are each 
$2^{N/2 -1}$-dimensional, and are known as the spinor
representations of the algebra $SO(N)$.
The entire kink spectrum for even $N$ consists of $2^{N/2}$
kinks in the two spinor representations. It was shown in \cite{meTBA}
that including these kinks along with the other particles
gives the correct free energy. This effectively proves that
these are all the kink states.

For odd $N=2P+1$, the spinor representation of dimension $2^P$ is
irreducible.  There is only one such spinor representation of
$SO(2P+1)$.  Thus it might seem that the simplest possibility is that
there are precisely $2^P$ kinks in the Gross-Neveu model for odd
$N$. However, Witten shows that there must be more \cite{Witten}. The
key is to look at the discrete symmetry $\psi^i \to -\psi^i$,
$\bar\psi^i \to -\bar\psi^i$.  One can think of this
symmetry as being $(-1)^F$, with the caveat that the fermion number
$F$ is not defined for $N$ odd: only $(-1)^F$ is.  This symmetry is
part of the $O(N)$ group, but not part of the connected
$SO(N)$ subgroup when $N$ is
odd. Because of the form (\ref{semi}), $b^i \to -b^i$ under this
symmetry. Thus there is a symmetry operator $\Lambda=(-1)^F$ which obeys 
$\Lambda^2=1$
and $\Lambda b^i \Lambda = -b^i$. 
In other words, $\Lambda$ anticommutes with all the
Clifford algebra generators, and commutes with the $SO(2P+1)$
generators. Thus the kink states in an irreducible representation of
$SO(2P+1)$ must all have the same eigenvalue of $\Lambda$. The fact that the
matrix element (\ref{matrixel}) does not vanish means that some kink
states have eigenvalue $+1$, while some have eigenvalue $-1$. Thus
there {\bf must} be more kink states than those in the single
irreducible representation of $SO(2P+1)$. Note that for even $N$, we
can identify $\Lambda$ with $\gamma^{N+1}$, and the multiple representations
with the two different spinor representations.  For odd $N$, there is
only one kind of spinor representation, so it must appear at least
twice.

So what happens for odd $N$? The simplest possibility is that there
are kinks in two spinor representations, each of dimension $2^P$
\cite{Witten,Wittlectures}. We
will see that this is essentially correct, but there is a major
subtlety. The key to the answer is in the generalized supersymmetry.

It is very useful to first examine the case $N=3$, where the answer is
known \cite{ZRSOS,Tsvelik,Schoutens,FI}. Here the Gross-Neveu model is
equivalent to the ${\cal N}=1$ supersymmetric sine-Gordon model
\cite{Witten} at a fixed value of the sine-Gordon coupling $\beta_{SG}$
where the model has an extra $SO(3)$ symmetry.
One might expect that the kink spectrum would generalize the kink and
and antikink of the ordinary sine-Gordon model to account for the
supersymmetry.  For the ${\cal N}=1$ supersymmetric sine-Gordon model,
it seems plausible that the kink and antikink each are a boson/fermion
doublet, making four particles in all \cite{SWsusy}.  However, this is
not the correct spectrum. The reason is the central charge $Z$
appearing in the supersymmetry algebra ((\ref{central}) with
$N=3$). For the supersymmetric sine-Gordon model, $Z$ does not vanish
\cite{WO}: it is the integral of a total derivative. $Z$ acting on
the kink states is non-zero because here the boundary conditions on
the field at positive spatial infinity and negative spatial infinity
are different.

The consequences of a non-vanishing $Z$ are familiar in supersymmetric
field theories. Representations where the mass $m=|Z|$ are
called BPS states \cite{BPS}. BPS
representations usually have a smaller number of states
than those with $Z=0$ \cite{WO} and
so are often referred to as ``reduced multiplets''.
When the BPS condition $m=|Z|$ holds, one can find a combination
of the generators which annihilates all BPS states. For the generalized
supersymmetry, the condition is
\begin{equation}
\left(Q - \frac{m^{N-2}}{Z} 
e^{(N-2)\theta} \overline Q\right) |\alpha\rangle =0
\end{equation}
for all BPS states $|\alpha\rangle$.

We need now to introduce a precise definition of the ``number of kink
states'' $K$. We say there are $K$ kink states if the number of
$n$-particle kink states at large $n$ depends on $n$ as
$K^n$. Put another way, the
entropy per kink of a gas of non-interacting kinks is $\ln K$.
$K$ is usually an integer in field theory: e.g.\ if there are two
one-particle states, there are four two-particle states, eight
three-particle states, and so on. The kink structure need not be so
simple, though. For example, consider kinks in a potential
$V(\phi)=(\phi^2- 1)^2$. The potential has minima at $\phi=\pm 1$,
so there are two one-kink states, having $\phi(\infty)=\pm 1$ and
$\phi(-\infty)=\mp 1$. However, there are only two two-kink states.
In fact, there are only two kink states for any $n$. Thus the 
number of kink states in this example is $K=1$.
The example of most importance for this paper is kinks
in a potential like $V(\phi)=\phi^2(\phi^2-1)^2$. Here the minima
are $\phi=0,\pm 1$. If kinks are allowed to only interpolate between
adjacent minima (i.e.\ only between $0$ and $\pm 1$), 
then the number of $n$-kink states doubles every time $n$ is increased
by $2$. Therefore, $K=\sqrt{2}$ in this triple-well potential.

For the BPS representations of the generalized supersymmetry
(\ref{central}), $K=\sqrt{2}$. This was already known for ${\cal N}=1$
supersymmetry in two dimensions \cite{ZRSOS,Schoutens,Wittlectures}, 
the $N$=3 case
here. The BPS representations was initially given for the thermal
($\phi_{1,3}$) perturbation of the tricritical Ising model, which has
${\cal N}=1$ supersymmetry as well as a Landau-Ginzburg description in
terms of a field $\phi$ with three degenerate minima \cite{ZRSOS}. To
reconcile these two facts, Zamolodchikov showed how supersymmetry
acts non-locally on the $\sqrt{2}$ kinks in
this triple-well potential.
Somewhat more surprisingly, there are $\sqrt{2}$ kinks in the
${\cal N}=1$ sine-Gordon model as well.

To give explicitly the action of the generalized supersymmetry
for odd $N$,
we denote a kink with $\phi(x=-\infty) = r$
and $\phi(x=+\infty)=s$ as $|rs\rangle$. In the triple-well
potential, $r,s=0,\pm$,
and $r-s=\pm 1$. The central charge $Z$ acting
on a kink of mass $m$ in a BPS representation is $m^{N-2}(r^2-s^2)$.
The BPS representation of the generalized supersymmetry for
odd $N$ is non-local:
it changes the kink configuration all the way to spatial $-\infty$.
On a single-particle state of rapidity $\theta$,
\begin{eqnarray*}
Q|r\,0\rangle = ir(me^\theta)^{(N-2)/2} |-r\,0\rangle\qquad&&\qquad
Q|0\,r\rangle = r(me^\theta)^{(N-2)/2} |0\,r\rangle\\
\overline Q|r\,0\rangle = ir(me^{-\theta})^{(N-2)/2} |-r\,0\rangle\qquad&&\qquad
\overline Q|0\,r\rangle = -r(me^{-\theta})^{(N-2)/2} |0\,r\rangle\\
\end{eqnarray*}
One can indeed verify that $Q$ and $\overline Q$ acting on this representation
satisfy the algebra (\ref{central}). Moreover, the discrete symmetry
$\Lambda$ acts non-diagonally as 
$$\Lambda|r\,s\rangle = |-r\,-s\rangle.$$
$\Lambda$ indeed anticommutes with the generalized supersymmetry generators,
With the interpretation of $\Lambda$ as $(-1)^F$, we see that
the kinks are neither bosons nor fermions.
Since the action of $Q$ and $\overline Q$ is non-local, we must be careful
to define their action on multiparticle states as well. For
example on a three-kink state, $Q$ acts as
$$Q(|\alpha\rangle\otimes |\beta\rangle\otimes |\gamma\rangle) =
(Q|\alpha\rangle)\otimes |\beta\rangle\otimes |\gamma\rangle
+ (\Lambda|\alpha\rangle)\otimes(Q|\beta\rangle)\otimes |\gamma\rangle +
(\Lambda|\alpha\rangle)\otimes(\Lambda|\beta\rangle)\otimes(Q|\gamma\rangle)$$
and likewise for $\overline Q$. Thus when $Q$ acts on a kink, it flips
all kinks to the left of it. For $N=3$, this reduces to the action
of supersymmetry discussed in \cite{ZRSOS,Schoutens}.

Of course, the kinks in the Gross-Neveu model must still transform as
a spinor of $SO(N)$. Thus we see that for odd $N$, the simplest
non-trivial kink spectrum consistent with all the symmetries of the
theory is for the kinks to be spinors of $SO(2P+1)$ {\bf and} interpolate
between the three degenerate wells of this potential.  Each kink is
labelled by a spinor index $1\dots 2^P$ and by a pair of vacua. The
number of kinks is $\sqrt{2}\times 2^P = 2^{N/2}$.  We emphasize that
the generalized supersymmetry and the discrete symmetry $\Lambda$ require
that there be more than the spinor: $2^{N/2}$ is the minimum number
of kinks satisfying these constraints.  In subsequent sections, we will
find the $S$ matrix for these kinks, and show that it gives the
correct free energy, effectively confirming the presence of these BPS
kinks.

\section{The $S$ matrices}

In this section we work out the $S$ matrix for the
kinks in the Gross-Neveu model for odd $N=2P+1$.

This $S$ matrix must obey a variety of
constraints.  These are easiest to write in terms of rapidity
variables, defined so that
the rapidity $\theta_1$ of a particle is related to its energy and
momentum by $E=m\cosh\theta_1$, $p=m\cosh\theta_1$. Lorentz invariance
requires that the two-particle $S$ matrix depend only on the
difference $\theta=\theta_1-\theta_2$. 
Any $S$ matrix arising from a unitary field theory must be
unitary and crossing-symmetric.
 Moreover, the integrability of the 
Gross-Neveu model means that the multi-particle $S$ matrix must
factorize into the product of two-particle ones. The resulting
constraint is called the Yang-Baxter equation.
Finally, the $S$ matrix must
obey the bootstrap equations. These mean that the $S$ matrix elements
of a bound state can be expressed in terms of the $S$ matrix of the
constituents. This constraint is explained in detail in \cite{OGW},
for example. 
Bound states show up as poles
in the $S$ matrix in the ``physical strip''
$0<Im(\theta)<\pi$.
If there is such a pole at $\theta=\theta_j$
in the kink-kink $S$ matrix, then the kinks of mass $m$
have a bound state at mass
\begin{equation}
m_{j}= 2m\cosh(\theta_j/2)
\label{bound}
\end{equation}

When $N=3$ and $N=4$, the kinks are the only states in the spectrum
\cite{Zunpub}.  For $N=3$, this was confirmed by the computation of
the exact free energy of the ${\cal N}=1$ supersymmetric sine-Gordon
model \cite{FI}.  For the latter, it follows because the $N=4$
Gross-Neveu model can be mapped to two decoupled ordinary sine-Gordon
models, each at the $SU(2)$-invariant coupling ($\beta_{SG}^2\to 8\pi$ in
the conventional normalization) \cite{Witten}.  This special
decoupling happens because the algebra $SO(4)$ is equivalent to
$SU(2)\times SU(2)$. At the $SU(2)$ invariant coupling of sine-Gordon,
there are no bound states in the spectrum: the only particles are the
kinks in doublets of each $SU(2)$.  In the $SO(4)$ language, these are
the spinor representations.  

For $N>4$, there are states other than the kinks. They have masses \cite{ZZ}
\begin{equation}
m_j = 2m\,{\sin\left(\frac{\pi j}{N-2}\right)}
\label{spectrum}
\end{equation}
where $m$ is the mass of the kink and $j=1\dots int(N-3)/2$. 
The first of these states
corresponds to the particle created by $\psi^i$; it is in the
$N$-dimensional vector representation of $SO(N)$. The other states are
bound states of the fermions.  One useful fact to note is that each
type of particle corresponds to a node on the Dynkin diagram for
$SO(N)$, as displayed in figure 1. For even $N$, the kinks correspond
to the two nodes on the right, while for odd $N$, the kink is the node
on the right. The vector representation is the node on the left.
\begin{figure}
\centerline{\epsfxsize=4.0in\epsffile{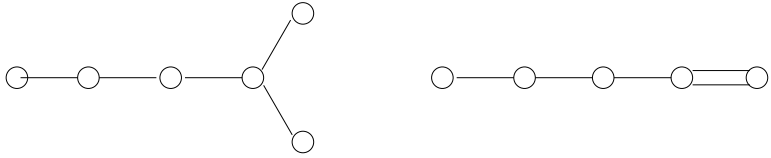}}
\bigskip\bigskip
\caption{The $SO(N)$ Dynkin diagrams for even $N$ and odd $N$}
\end{figure}

The exact $S$ matrix of the vector particles for
any $N$ was worked out in \cite{ZZ}. Using the bootstrap gives
the $S$ matrix for fermion bound states, but not the kinks.
The explicit expression for the $S$ matrix of
particles with mass $m_2$ (which
are in the antisymmetric and singlet representations of $SO(N)$)
can be found in \cite{Mackay}. The $S$ matrix for the kinks
for even $N$ is worked out in  \cite{SW,KT} (see also \cite{OGW}).
Thus to complete this picture we need to work out the scattering
of the kinks for odd $N$, including the fact that they have the
additional BPS structure discussed in the last section.

The BPS kinks in the Gross-Neveu model are in the $2^P$-dimensional
spinor representation of $SO(2P+1)$, and they also form a multiplet
transforming under the generalized supersymmetry.
Since the two symmetries
commute with each other, the simplest two-particle
$S$ matrix invariant under these symmetries
is of the tensor-product form
\begin{equation}
S(\theta)= S_{spinor}(\theta) \otimes S_{BPS}(\theta).
\label{tensor}
\end{equation}
The matrices $S_{spinor}(\theta)$ and
$S_{BPS}(\theta)$ are respectively the $S$ matrices
for particles in the spinor representation of $SO(2P+1)$, and for kinks
in the triple-well potential.

The spinor part of the $S$ matrix was found in \cite{OGW}.  The fact
that it is invariant under $SO(2P+1)$ means that it can be written in
terms of projection operators. A projection operator ${\cal P}_a$ maps
the tensor product of two spinor representations onto an irreducible
representation labelled by $a$. Here $a=0\dots P$, where $a=0$ labels
the identity representation, $a=1$ the vector representation, $a=2$
the antisymmetric tensor, and so on up to $a=P-1$. The representation
with $a=P$ is the representation with highest weight $2\mu_P$, where
$\mu_P$ is the highest weight of the spinor representation.  We
do not give the explicit expressions for the projectors because we
will not need them; they can be written explicitly in terms of the gamma
matrices for the algebra $SO(2P+1)$, which in turn can be written as
tensor products of Pauli matrices.  The spinor $S$ matrix is
\begin{equation}
S_{spinor} (\theta) = \sum_{a=0}^{P} f_a(\theta) {\cal P}_a (\theta)
\label{spinor}
\end{equation}
The functions $f_a(\theta)$ are not constrained by the
$SO(2P+1)$ symmetry, but all can be related to $f_P(\theta)$
by using the Yang-Baxter equation.  The result is \cite{OGW}
\begin{eqnarray}
\nonumber
f_{P-1}&=&\frac{\theta + i\pi\Delta}{\theta - i\pi\Delta}\ f_P\\
f_{P-a-1} &=&\frac{\theta+i\pi\Delta(2a+1)}
{\theta-i\pi \Delta(2a+1)}\ f_{P-a+1}
\label{recursion}
\end{eqnarray}
where the Yang-Baxter equation does not determine $\Delta$.

The function $f_P(\theta)$ must satisfy
the unitarity and crossing relations. These do not determine the
function uniquely, because one can obtain a new solution of these
equations by multiplying any given solution by a function $F(\theta)$
obeying $F(\theta)=1$ and $F(\theta)=F(i\pi-\theta)$. This is known as
the CDD ambiguity. The minimal solution of the unitarity and crossing
relations is a solution without any poles in the physical strip.
The minimal spinor $S$ matrix of $SO(2P+1)$ has 
$$\Delta =\frac{1}{N-2}= \frac{1}{2P-1}$$ 
and \cite{OGW}
\begin{equation}
f_P(\theta) =  \prod_{b=0}^{P-1}
\frac{\Gamma\left(1-b\Delta -\frac{\theta}{2\pi i}\right)
\Gamma\left( (b+\frac{1}{2})\Delta +  \frac{\theta}{2\pi i}\right)}
{\Gamma\left(1-b\Delta +\frac{\theta}{2\pi i}\right)
\Gamma\left( (b+\frac{1}{2})\Delta -  \frac{\theta}{2\pi i}\right)}
\label{fp}
\end{equation}
Notice that this has zeroes at $\theta=i\pi \Delta(2b+1)$ for all
$b=0\dots P-1$. These cancel poles arising from (\ref{recursion}),
and ensure that there are no poles in the physical
strip any of the $f_a$. 
The following integral representation for $f_P$ will be useful in the future:
\begin{equation}
f_P(\theta)=\exp\left[\int_{-\infty}^\infty {d\omega\over\omega} e^{i(2P-1)\omega\theta/\pi} 
{e^{-|\omega|/2}\sinh (P\omega)\over 2\cosh [(2P-1)\omega/2] \sinh(\omega)}\right]
\end{equation}

The minimal $S$ matrix for BPS kinks in a triple-well potential can
also be determined by imposing the same criteria.  In addition, we can
utilize the generalized supersymmetry.  Because the generalized
supersymmetry operators $Q$ and $\overline Q$ commute with the Hamiltonian,
they must also commute with the $S$ matrix.  Since we know how they
act on multiparticle states, this is simple to implement. In fact,
some of the work has already been done for us. 
The $S$ matrix for the tricritical Ising model describes
the scattering of kinks in a triple well, and is invariant
under ordinary supersymmetry \cite{ZRSOS}.
The generalized supersymmetry algebra acting on the
states (\ref{central}) is related to the ordinary supersymmetry
algebra by making the substitution $\theta\to \pm (N-2)\theta$. 
Thus in order to commute with the generalized supersymmetry,
$$ S_{BPS}(\theta) \propto S_{TCI}((N-2)\theta) \qquad\hbox{or}\qquad
S_{BPS}(\theta) \propto S_{TCI}(-(N-2)\theta) .$$ Such an $S$ matrix
will automatically satisfy the Yang-Baxter equation as well (any
solution remains a solution under the scaling $\theta\to\lambda\theta$).

\begin{figure}
\centerline{\epsfxsize=1.0in\epsffile{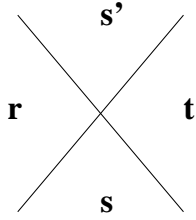}}
\bigskip\bigskip
\caption{Representing kink scattering by four vacua}
\end{figure}
It is easiest to label the $S$ matrix elements for kinks by their
vacua.  A two-kink configuration can be labeled by three vacua. As
shown in figure 2, a two-particle $S$-matrix element can be labeled by
four vacua because only the middle vacuum can change in a
collision. Thus this $S$ matrix element $S^{(rt)}_{ss'}$ describes
scattering the initial state $rst$ to the final state $rs't$.  For the
triple well, the labels $r,s,s',t$ take the values $0,\pm 1$. The
elements of $S_{BPS}$ are then
\begin{eqnarray}
\nonumber
S^{(r-1\,r+1)}_{rr} (\theta) &=& B(\theta)
\left({\beta_r
\over\beta_{r+1}^{1/2}\beta_{r-1}^{1/2}}\right)^{i{\theta\over\pi}}
i\sinh\left[\lambda\theta-\frac{i\pi}{4}\right]\\
\nonumber
S^{(rr)}_{r\pm1\, r\mp1}(\theta) &=&B(\theta)\left(
{\beta_{r+1}^{1/2}\beta_{r-1}^{1/2}\over \beta_r}\right)^{1+i{\theta\over\pi}}
(-1)^P i\sinh[\lambda\theta] \\
\nonumber
S^{(rr)}_{r+1\,r+1}(\theta) &=&
B(\theta)\left({\beta_{r+1}\over \beta_r}\right)^{i{\theta\over\pi}}
{\beta_1\over \beta_r}\cosh\left[\frac{i\pi}{4} r+\lambda\theta\right]\\
S^{(rr)}_{r-1\,r-1}(\theta) &=& B(\theta)
\left({\beta_{r-1}\over \beta_{r}}\right)^{i{\theta\over\pi}}
{\beta_1\over \beta_r}\cosh\left[\frac{i\pi}{4} r-\lambda\theta\right]
\label{smat}
\end{eqnarray}
where 
$$\beta_r = \cos(\frac{\pi}{4} r).$$
The tricritical Ising model $S$ matrix has $\lambda = 1/4$,
so in general $\lambda$ must be $\pm (N-2)/4$.
Crossing symmetry fixes the sign, requiring that
$$\lambda = (-1)^{P+1} \frac{N-2}{4} = (-1)^{P+1} \frac{1}{4\Delta}$$
We should note that some extra minus signs appear in the crossing
relations because the kinks are neither bosons nor fermions \cite{KT}.
This solution of the Yang-Baxter equation was originally found in an
integrable lattice model, the hard-hexagon model, where
the kink vacua correspond to the heights in the lattice model. This
lattice model is part of the ``restricted solid-on-solid'' (RSOS)
hierarchy \cite{ABF}. More general RSOS models
let the heights/vacua run over more values; these arise in the multi-flavor
and symplectic Gross-Neveu models to be discussed in \cite{us}.

The last step is to find the function $B(\theta)$.
Crossing symmetry requires that
$$B(\theta) = B(i\pi-\theta)$$
while unitarity requires that
$$B(\theta)B(-\theta)= 
\frac{1}{\sinh\left[\lambda\theta-\frac{i\pi}{4}\right] 
\sinh\left[\frac{i\pi}{4} + \lambda\theta\right] }.
$$
One can easily find the minimal solution of these relations,
where $B(\theta)$ has no poles in the physical strip
$0<Im\theta<\pi$. 
For the $N=3$ case,
the ${\cal N}=1$ supersymmetric sine-Gordon model,
there are no bound states: the BPS kinks make up the entire spectrum
\cite{Witten}.
There are no poles in the physical strip for $N=3$.
This result is confirmed by calculating the free energy \cite{FI},
a computation we will repeat in the next section.

In general, we need a $B(\theta)$ with poles, because the Gross-Neveu
model has bound states.  For $P>1$, there are particles other than the
kink in the spectrum.  These have masses given by (\ref{spectrum}),
with $j=1\dots P-1$.  The fermion $\psi_i$ is the $j=1$ state, and
higher $j$ corresponds to bound states of the fermion \cite{ZZ}.  For
odd $N=2P+1$ all fundamental representations except for the spinor
appear in the tensor product of two spinor representations.  Moreover,
the poles in (\ref{recursion}) correspond to bound-state particles in
the correct representations of $SO(2P+1)$.  We thus expect all states
(other than the kink itself) to appear as bound states of two kinks.
It follows from (\ref{bound}) and (\ref{spectrum}) that the bound
states appear as poles in $B(\theta)$ at $\theta_j=i\pi(1-2j\Delta)$
for $j=1\dots P-1$. Crossing symmetry means that there must be
poles at $\theta_j=i\pi 2j\Delta$ as well.

We find that
\begin{eqnarray*}
B(\theta)&=& \frac{i}{\sinh\left[\lambda\theta+\frac{i\pi}{4}\right]}
\prod_{b=1}^{int\{(P-1)/2\}} \frac{
\sinh\left[\frac{1}{2}(\theta + i\pi - i 4b \pi\Delta)\right]
\sinh\left[\frac{1}{2}(\theta + i 4b\pi \Delta)\right]}
{\sinh\left[\frac{1}{2}(\theta - i\pi + i 4b \pi\Delta)\right]
\sinh\left[\frac{1}{2}(\theta - i 4b \pi \Delta)\right]}\\
&&\quad \times \prod_{l=1}^\infty \frac{
\Gamma\left[\frac{1}{2}+ (l-1)\frac{1}{2\Delta} - \frac{\theta}{4\pi i\Delta}\right]
\Gamma^2\left[\frac{1}{2}+ (2l-1)\frac{1}{4\Delta} +\frac{\theta}{4\pi i\Delta}\right]
\Gamma\left[\frac{1}{2}+ l\frac{1}{2\Delta} - \frac{\theta}{4\pi i\Delta}\right]}
{\Gamma\left[\frac{1}{2}+ (l-1)\frac{1}{2\Delta} +\frac{\theta}{4\pi i\Delta}\right]
\Gamma^2\left[\frac{1}{2}+ (2l-1)\frac{1}{4\Delta} - \frac{\theta}{4\pi i\Delta}\right]
\Gamma\left[\frac{1}{2}+ l\frac{1}{2\Delta} +\frac{\theta}{4\pi i\Delta}\right]}.
\end{eqnarray*}
The $1/\sinh(\lambda\theta + i\pi/4)$ gives poles corresponding to
bound states with odd $j$. The finite product of $\sinh$ functions
gives the appropriate poles for even $j$. The infinite product
of gamma functions assures that crossing symmetry is obeyed.
Another way of writing $B(\theta)$ is as
\begin{eqnarray*}
B(\theta)= \frac{i}{\sinh\left[\frac{i\pi}{4}-\lambda\theta\right]}
\prod_{b=1}^{int\{P/2\}} \frac{
\sinh\left[\frac{1}{2}(\theta + i\pi - i (4b-2) \pi\Delta)\right]
\sinh\left[\frac{1}{2}(\theta + i (4b-2) \pi\Delta)\right]}
{\sinh\left[\frac{1}{2}(\theta - i\pi + i (4b-2) \pi\Delta)\right]
\sinh\left[\frac{1}{2}(\theta - i (4b-2) \pi\Delta)\right]}\\
\times \prod_{l=1}^\infty \frac{
\Gamma\left[1+ (l-1)\frac{1}{2\Delta} - \frac{\theta}{4\pi i\Delta}\right]
\Gamma\left[ (2l-1)\frac{1}{4\Delta} +\frac{\theta}{4\pi i\Delta}\right]
\Gamma\left[1+ (2l-1)\frac{1}{4\Delta} +\frac{\theta}{4\pi i\Delta}\right]
\Gamma\left[ l\frac{1}{2\Delta} - \frac{\theta}{4\pi i\Delta}\right]}
{\Gamma\left[1+ (l-1)\frac{1}{2\Delta} +\frac{\theta}{4\pi i\Delta}\right]
\Gamma\left[ (2l-1)\frac{1}{4\Delta} - \frac{\theta}{4\pi i\Delta}\right]
\Gamma\left[1+ (2l-1)\frac{1}{4\Delta} - \frac{\theta}{4\pi i\Delta}\right]
\Gamma\left[ l\frac{1}{2\Delta} +\frac{\theta}{4\pi i\Delta}\right]}.
\end{eqnarray*}
Here the $1/\sinh(\lambda\theta - i\pi/4)$ gives poles corresponding to
bound states with even $j$, while the finite product of $\sinh$ functions
gives the appropriate poles for odd $j$. With a little tedium one
can show that the two expressions for $B(\theta)$ are in fact
equal. 
The following integral representation will be useful: 
\begin{equation}
B(\theta)={\sqrt{2}\over \sqrt{\cosh
    2\lambda\theta}}\exp\left[-\int_{-\infty}^\infty
{d\omega\over \omega} e^{i(2P-1)\omega\theta/\pi} ~~
{2\sinh [(P-3/2)\omega]+\sinh[(P-1/2)\omega]\over
4\sinh[\omega]\cosh[(P-1/2)\omega)]}\right]
\end{equation}
It is most easily obtained by multiplying the two expressions for
$B(\theta)$, a standard rewriting of the logarithm of the gamma
functions in terms of an integral, and then taking the square root.

We have thus determined the $S$ matrix (\ref{tensor}) for the
BPS kinks in the Gross-Neveu model.  

By using the bootstrap, one
obtains the $S$ matrix for all the particles, for example reproducing
the $S$ matrix of \cite{ZZ} for the fermions $\psi_i$.  
To apply the bootstrap, we need to utilize a number of results from
integrable lattice models. In this context, the process of obtaining a
new solution of the Yang-Baxter equation from an existing one is
called fusion \cite{KRS,Kyoto}. For example, in a model with $SU(2)$
symmetry, one can fuse the $S$ matrices for two spin-$1/2$ particles
to get that for a spin-$1$ particle.  In the simplest cases, the
fusion procedure utilizes the fact at a pole $\theta=\theta_j$ of the
$S$ matrix, the residue $(\theta-\theta_j)S(\theta_0)$ becomes a
projection operator. This happens only at imaginary values of the
rapidity, so this does not violate unitarity.  For example, from
(\ref{recursion}), it follows that at $\theta=\theta_1\equiv
i\pi(1-2\Delta)$, all the $f_j(\theta_1)=0$ except for $j=1$.  Thus
$S_{spinor}$ becomes a projection operator ${\cal P}_1$ onto the $j=1$
bound states. This means that one can think of the fermion of
rapidity $\theta$ as a bound
states of two kinks, one with rapidity $\theta+\theta_1/2$ and
the other with $\theta-\theta_1/2$.
In general, consider a case where the $S$ matrix of two particles
$\alpha$ and $\beta$ has a pole at $\theta=\theta_{j}$. 
The residue of the $S$ matrix at the pole is the sum of projection operators
$$\lim_{\theta\to\theta_j} \left[(\theta-\theta_j)
S^{(\alpha\beta)}(\theta)\right] = 
\sum_a R_a {\cal P}_a$$
The $S$ matrix of the bound state $j$ from another particle $\gamma$
related to the $S$ matrices of the constituents by the formula
\cite{KRS,OGW}
\begin{eqnarray}
S^{(j\gamma)}=\left(\sum_a \sqrt{|R_a|} {\cal P}_a\right)  
S^{(\alpha\gamma)}(\theta+\theta_{j}/{2})
S^{(\beta\gamma)}(\theta-\theta_{j}/{2})
\left(\sum_a\frac{1}{\sqrt{|R_a|}}{\cal P}_a\right).
\label{bootstrap}
\end{eqnarray}
Note that the matrices in (\ref{bootstrap}) are not all acting on the
same spaces, so this relation is to be understood as multiplying the
appropriate elements.  The upshot is that
one can think of the particle
with mass $m_j$ and rapidity $\theta$ as the bound state of kinks with
rapidities $\theta+\theta_j/2$ and $\theta-\theta_j/2$.

The $f_a$ given by (\ref{recursion}) determine in which
representation of $SO(2P+1)$ the particles for a given $j$
transform. For example, the particles with mass $m_{1}$ transform in
the vector representation, so there are $2P+1$ of them. These are the
original fermions. The particles with mass $m_2$ transform in the
antisymmetric representation and the singlet, so there are $2P^2+P+1$
of these.  In general, the particles with mass $m_j$ are in the
$a$-index antisymmetric tensor representations with $a=j,j-2,
\dots$. The reason for the
multiple representations of particles at a given mass is likely that they
form an irreducible representation of the Yangian algebra for $SO(2P+1)$
\cite{Mackay}.
To compute the explicit
$S$ matrix for all the bound states is a formidable task; for some
results, see \cite{Mackay}.
However, by using fusion we can compute the ``prefactor'' of the $S$ matrix,
which is necessary for the computation of the free energy in the next
section.
The prefactor of the $S$ matrix for scattering a particle in 
representation with highest weight $\mu_a$ from
a particle in representation $\mu_b$ is defined as
the $S$ matrix element multiplying the projector on the representation
with highest weight $\mu_a+\mu_b$. We denote the prefactor
as ${\cal S}_{ab}(\theta)$, and for example, for kink-kink scattering
the prefactor is
$${\cal S}_{PP}(\theta) = B(\theta) f_P(\theta).$$

An important check on this $S$ matrix is that the scattering
closes. This means that all the poles in the physical strip correspond
to particles in the spectrum. In fact, zeroes coming from the BPS part
of the $S$ matrix cancel spurious poles coming from the spinor part:
without the BPS part, the bootstrap does not close. This was noted in
\cite{OGW}, where it was termed a `violation'' of the bootstrap for
spinor particles \cite{OGW}. Solving this problem gives another way of
seeing that the extra BPS structure must be present.
To find the zeros coming from fusing $S_{BPS}$, note that $\theta_j$
for odd $j$ obeys $\sinh(\lambda\theta_j + i\pi/4)=0$,
while for even $j$ it obeys obey $\sinh(\lambda\theta_j - i\pi/4)=0.$
Thus at $\lambda\theta=-i\pi/4$, the $S$ matrix in (\ref{smat})
projects onto the states
$$ |-1\,1\rangle,\ |1\,-1\rangle,\ |0\, 0\rangle \qquad\quad j\hbox{ odd}$$
while at $\lambda\theta=i\pi/4$, it projects onto
$$|-1,\, -1\rangle,\ |0\, 0\rangle,\ |1\, 1\rangle \qquad\quad j\hbox{ even}.$$
Using the formula (\ref{bootstrap})
yields the bound-state $S$ matrix. 
One finds that fusing $S_{BPS}$ gives
an $S$ matrix effectively diagonal
in the kink labels, and moreover, it is the same for all labels.
Thus it contributes
only an overall factor to the $S$ matrix of a bound
state $j$ from a kink $\gamma$, namely
\begin{eqnarray*} 
S^{(j\gamma)}&\propto &\sinh\left[\lambda\theta + \frac{i\pi}{8}\right]
\sinh\left[\lambda\theta - \frac{i3\pi}{8}\right] \qquad j\hbox{ odd}\\
S^{(j\gamma)}&\propto &\sinh\left[\lambda\theta - \frac{i\pi}{8}\right]
\sinh\left[\lambda\theta + \frac{i3\pi}{8}\right] \qquad j\hbox{ even}
\end{eqnarray*}
This overall factor is the only effect of the restricted-kink structure
on the bound states $j=1\dots P-1$. The non-integer number of kink states
and the funny rules do not affect the bound states.
However, these extra factors do result in zeroes which cancel spurious
poles occurring in the fusion of $S_{spinor}$ for odd $N$.

There is an algebraic explanation for the simplicity of the kink
structure in the bound states.
Conceptually, it is similar to
the cases with a Lie-algebra symmetry, but here
it is because $S_{BPS}$ has a quantum-group symmetry. 
In our case of interest, the quantum-group
algebra is $U_q(SU(2))$, with the deformation parameter $q=e^{2\pi
i/3}$. This algebra is a deformation of $SU(2)$, and many of its
representations have properties similar to ordinary $SU(2)$. In
particular, the kinks interpolating between adjacent minima are the
spin-$1/2$ representations of $U_q(SU(2))$. However, when $q$ is a
root of unity, various special things happen. In our case of a triple
well, what happens is that the spin-$1$ representation is very simple.
A spin-$1$ kink goes from either the vacuum $1$ to $-1$, $-1$ to $1$,
or from $0$ to $0$. Note that whichever vacua one is in, there is
only one vacua to go to. Hence the number of spin-$1$ kinks is $K=1$.
The triple-well structure can be ignored once one fuses the kinks:
effectively the BPS structure goes away for the bound states.

To conclude this section, we note that kinks transforming under
the generalized supersymmetry have appeared before, 
in the ordinary sine Gordon model at coupling $\beta_{SG}^2 =
16\pi/(2P+1)$  \cite{Itoyama}. In fact,  the Gross-Neveu model
and the sine-Gordon model at these couplings are closely
related. The masses $m_j$ (but not the degeneracies) in the
spectrum are identical, with the soliton and antisoliton of the
sine-Gordon model having the mass $m$ like the kinks here
\cite{ZZ}. Even more strikingly, the soliton and antisoliton are
representations of the generalized supersymmetry (\ref{central}), as
shown in \cite{Itoyama}. The reason for the similarity is
that both are related to a certain integrable perturbation 
(by the (1,1;adjoint) operator) of the
coset conformal field theories
$$\frac{SO(N)_k \times SO(N)_1}{SO(N)_{k+1}}$$
The $k=1$ case is the sine-Gordon model at $\beta_{SG}^2=16\pi/N$,
while the $k\to\infty$ case is the Gross-Neveu model \cite{mesigmacoset}.
The models for all $k$ should be integrable, and for
odd $N$ the $S$ matrix should
have the form
$$S= S_{q}\otimes S_{BPS}$$
where $q$ is related to $k$.
The $S$ matrix $S_{q}$ is the $S$ matrix associated with the spinor
representation of the quantum group $U_q(SO(N))$, which as far as we know
has never been worked out explicitly in general. 
In the $k=1$ sine-Gordon case, the
generalized supersymmetry is extended to an ${\cal N}$=2 version, with
generators $Q_\pm,\overline{Q}_\pm$. This means that for $k=1$,
$S_q$ is in fact $\propto S_{BPS}$. 
For $k\to\infty$, $q\to 1$
and the quantum-group algebra reduces to the usual Lie algebra $SO(N)$.
Thus $\lim_{q\to 1} S_{q} = S_{spinor}$.

\section{The free energy}

In this section we compute the free energy of the Gross-Neveu model at
finite temperature using a technique called the thermodynamic Bethe
ansatz (TBA). This gives a
non-trivial check on the exact $S$ matrix.  In the limit of all masses
going to zero, the theorem of \cite{BCNA} says that the free energy
per unit length must behave as
\begin{equation}
\lim_{m\to 0} F = -\frac{\pi T^2}{6} c_{UV}
\label{freeuv}
\end{equation}
where $c_{UV}$ is the central charge of the conformal field theory
describing this ultraviolet limit. The number $c_{UV}$ can usually be
calculated analytically from the TBA, because in this limit the free
energy can be expressed as a sum of dilogarithms.  The $c_{UV}$
computed from the TBA must of course match the $c_{UV}$ from the field
theory. This provides a very useful check, because the
spectrum is essentially an infrared property. Finding the central
charge exactly in the ultraviolet limit requires knowing not only the
exact spectrum but the exact $S$ matrix as well: the gas of particles
at high temperature is strongly interacting. All particles contribute
to the free energy, so if some piece of the spectrum is missing, or if
an $S$ matrix is wrong, the TBA will not give the correct $c_{UV}$.

The free energy is computed from the TBA by two steps. First one finds
how the momenta are quantized when periodic boundary conditions are imposed.
Precisely, one demands that the multi-particle wave function
obeys $\psi(x_1,x_2,\dots)$ remain the same when any coordinate
$x_i$ is shifted to $x_i +L$. In an interacting theory, the quantization
involves the $S$
matrix, because as one particle is brought around the periodic world,
it scatters through the other particles. In the continuum, this
leads to a constraining relation between the
densities of states and the actual particle densities. The free energy
at temperature $T$
is found in the second step by minimizing it subject to the
constraint. The detailed procedure for this computation has been
discussed in many places, so we will not repeat these
explanations. Several papers closely related to the current
computation are \cite{Alyosha,FI,meTBA}.  The TBA computation here is
technically complicated because the scattering is not diagonal. This
means that, as a particle is going around the periodic interval of
length $L$, it can change states as it
scatters through the other particles.
The way to proceed then is well known. First, one has to set up the
system of auxiliary Bethe equations to diagonalize the transfer
matrix, then use its results to determine the allowed rapidities. 
This computation amounts to a standard Bethe ansatz computation.
The end result can be written conveniently by introducing extra
zero-mass ``pseudoparticles'' or ``magnons''
to the constraining relations.
Then one minimizes the free energy to find out the equilibrium
distributions at temperature $T$, and thus the thermal properties of
the 1+1-dimensional quantum field theory. 

Two kinds of additional complications occur here.  The first
difficulty is that, as a striking consequence of the bootstrap
analysis, particles appear in
all $j$-index antisymmetric tensor representations $\Pi_j$, 
but in general there is not a single mass associated with a given
representation. Instead, particles with mass $m_a$ appear in
representations $j=a,a-2,\ldots$, a mixture we call
$\rho_a$. On the other hand, little is known about the diagonalization
of transfer matrices acting on products of $SO(2P+1)$
representations. To proceed, we have to assume that the usual Bethe
equations based on the Dynkin diagram of the underlying Lie algebra
\cite{Reshetikhin} apply, in the $SO(2P+1)$ case, not to the
irreducible representations $\Pi_j$, but precisely to the mixtures
$\rho_a$. This is a key technical assumption, which is checked a
posteriori, because the number of particles of each mass can be read
off from the TBA equations.

The second is that the kinks in the Gross Neveu model are not only in
the $2^P$-dimensional spinor representation, they also form a
multiplet under the generalized supersymmetry transformations. This
means that instead of one, one gets two auxiliary problems, one to
diagonalize the transfer matrices acting on the degrees of freedom
transforming in the spinor representation, the other to diagonalize
the transfer matrices acting on the BPS degrees of freedom.  A similar
but simpler problem has been solved in the TBA calculation of the
${\cal N}=1$ supersymmetric sine-Gordon model \cite{FI}.

We first solve the auxiliary problem for the BPS $S$ matrix. This
was already done in \cite{Alyosha}, because $S_{BPS}$ is the
same as that in the tricritical Ising model, up to a rescaling of $\theta$.
The result is that the TBA equations in the tricritical
Ising model require introducing one pseudoparticle with density
$\tau(\theta)$, and hole density $\tilde{\tau}(\theta)$.
Then the density of states for the particle in the tricritical Ising
model is
$$2\pi P_{TCI}(\theta) = m\cosh\theta + {1\over 2} {\xi\star\xi\over 2\pi}\star
\rho_{TCI} +{1\over 2} \xi\star(\tau-\tilde{\tau})$$
where we have
defined convolution as
$$a\star b(\theta) \equiv \int_{-\infty}^\infty a(\theta-\theta') b(\theta') 
d\theta'.$$
The kernel $\xi(\theta)$ is defined as
\begin{equation}
\xi(\theta)=\frac{2P-1}{\cosh[(2P-1)\theta]}
\end{equation}
in general, with $P=1$ for the tricritical Ising model.
The density of real particles is denoted by $\rho_{TCI}$. 
It is most convenient to give most kernels
in terms of Fourier transforms, defined as
$$
\hat{f}(\omega)\equiv \int {d\theta\over 2\pi} e^{i(2P-1)
\omega\theta/\pi} f(\theta).$$
so that $\hat{\xi}(\omega)={1\over 2\cosh({\omega\over 2})}$.

The density of states for the pseudoparticles is then 
related to the density of real
particles $\rho_{TCI}$ by
$$2\pi\left(\tau+\tilde\tau\right)=\xi\star \rho_{TCI}$$ The
pseudoparticles have zero energy, but they contribute entropy to the
free energy. The equilibrium values $\rho_{TCI}$ and $\tau$ are found
by minimizing the free energy subject to the above constraints.
Notice that simple manipulations allow us to reexpress the density of
states for the particle as 
$$
2\pi P_{TCI}(\theta)=m\cosh\theta+{\xi\star \xi\over 2\pi}\star
\rho_{TCI}-\xi\star\tilde{\tau}=m\cosh\theta+\xi\star\tau
$$
The second form will be the most convenient in what follows.

Before giving the full answer for the odd-$N$ Gross-Neveu model,
let us first also  review the solution for the $P=1$ case, the ${\cal N}=1$
supersymmetric sine-Gordon model at its $SO(3)$ symmetric point. As discussed
in the last section, the $S$ matrix is
$$S_{SO(3)} = S_{TCI} \otimes S_{\beta_{SG}^2\to 8\pi}.$$ The latter piece
is the $S$ matrix for the ordinary sine-Gordon model at an
$SU(2)$-invariant point ($\beta_{SG}^2\to 8\pi$ in the usual
normalization).  The auxiliary problem for this piece is equivalent to
diagonalizing the Heisenberg spin chain, a problem solved by Bethe 70
years ago. One must introduce an infinite number of pseudoparticles
with densities $\rho_j(\theta)$, in addition to the pseudoparticle 
density $\tau(\theta)$ required for the BPS $S$ matrix.
The density of states $P_0$ for the kinks of mass $m$ is then related
to the particle density $\rho_0$ given by
\begin{equation}
2\pi P_0 (\theta) = {m}\cosh\theta + 
{\cal Y}\star\rho_0(\theta) - \sum_{j=1}^\infty
\sigma^{(\infty)}_{j} \star \widetilde\rho_j(\theta)- \xi\star\tilde{\tau}(\theta).
\label{bet0}
\end{equation}
The kernels 
$\sigma^{(\infty)}_{j}$ follow from the Bethe ansatz analysis, and
are
$$\hat{\sigma}_j^{(\infty)}=e^{-j|\omega|}.$$ The kernel ${\cal Y}$
comes from two places. There is a contribution from the prefactor of
the $S$ matrix, and for kink-kink scattering there is an extra piece
arising from the Bethe ansatz analysis \cite{Alyosha}. The formula for
any $P$ is
\begin{equation}
\hat{\cal Y}_{PP}^{(P+1/2)}=\widehat{{d\over d\theta}\hbox{Im}\ln B}+ \widehat{
{d\over d\theta}\hbox{Im}\ln f_P} + {1\over 2}
 \left(\hat{\xi}\right)^2 
\label{Ypp}
\end{equation}
so that 
$$\hat{\cal Y}=\hat{\cal Y}^{(P+1/2)}_{PP}\big|_{P=1}=1-\frac{e^{|\omega|}}
{4\cosh^2(\omega/2)}$$
The other Bethe equations relate the densities of states for
the pseudoparticles to particle and pseudoparticle densities.
They are 
\begin{equation}
2\pi {\rho}_j(\theta) = 
\sigma_j^{(\infty)} \star \rho_0(\theta) - \sum_{l=1}^\infty
A_{jl}^{(\infty)} \star \widetilde{\rho}_l(\theta)
\label{betj}
\end{equation}
where the density of string states $P_j$ is
$$P_j = \widetilde{\rho}_j + \rho_j.$$
The new kernel is given by
$$\hat{A}_{jl}^{(\infty)}=2\coth
(\omega)~ e^{-max(j,l)|\omega|}~\sinh(min(j,l)\omega).$$ 
These equations can be simplified greatly by inverting the matrix $A_{jl}$
(for details see for example \cite{meTBA}), giving
\begin{equation}
2\pi {P}_j(\theta) = \delta_{j0} m \cosh\theta + \xi \star 
(\rho_{j-1}+\rho_{j-1})
\label{betall}
\end{equation}
Even the pseudoparticle $\tau$ coming from the BPS piece is included in
this equation, by defining $\rho_{-1}(\theta)\equiv\tau(\theta)$ and 
$P_{-1}=\tau+\tilde\tau$. Minimizing the free energy then yields the TBA
equations
$$\epsilon_j(\theta) = \delta_{j0} m\cosh\theta - \xi \star
\left[\ln (1+e^{-\epsilon_{j-1}/T}) + \ln (1+e^{-\epsilon_{j+1}/T})\right]$$
with $j=-1,0,1,2,\dots$, and $\epsilon_{-2}\equiv \infty$. These
TBA equations are conveniently encoded in the diagram in figure 3.
The circles represent the functions
$\epsilon_a$; the filled node represents the fact that the equation
for $\epsilon_0$ has a mass term.
\begin{figure}
\centerline{\epsfxsize=4.0in\epsffile{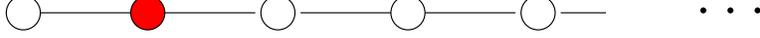}}
\bigskip\bigskip
\caption{The TBA for the $O(3)$ Gross-Neveu model (the 
supersymmetric sine-Gordon model).}
\end{figure}
The free energy per unit length $F$ is
\begin{equation}
F (m,T)=
 - T  m \int_{-\infty}^{\infty} \frac{d\theta}{2\pi} \cosh\theta
\ln\left( 1 + e^{-\epsilon_{0}(\theta)/T}\right)
\label{free}
\end{equation}
One can check that in the limit $m\to 0$, this free energy yields 
the correct central charge $3/2$ \cite{FI}. This confirms the presence
of the BPS kinks in the spectrum.

The TBA for the full $O(2P+1)$ Gross-Neveu model is conceptually
similar to the $O(3)$ case, but is much more involved technically. To
complicate matters further, there are particles with masses $m_a$,
$a=1\dots P$.  Luckily, many of the technical complications have
already been solved.  The diagonalization of the BPS part is the same
as in the tricritical Ising model \cite{Alyosha}.  This requires
introducing the pseudoparticle density $\tau(\theta)$, as described
above.  The diagonalization of the auxiliary problem for the $SO(2P+1)$
spinor part is done using the standard string hypothesis based on the
Dynkin diagram \cite{Kuniba}. For the TBA for the $SU(N)$ and $O(2P)$
Gross-Neveu models discussed in \cite{meTBA}, the TBA equations are
related to the $SU(N)$ and $SO(2P)$ Dynkin diagrams. 
However, $SO(2P+1)$ is
not simply laced, so this case is somewhat more complicated.
obvious. Nevertheless, the appropriate computation has been done
already in \cite{Kuniba}, utilizing the Yangian structure of the $S$
matrix. The densities of particles with mass
$m_a$ are defined as $\rho_{a,0}$, while the auxiliary problem
requires introducing pseudoparticle densities $\rho_{a,j}$ with
$j=1\dots\infty$, as well as $\tau$. In our notation, we try to remain
consistent with \cite{meTBA}, where the TBA for the $O(2P)$
Gross-Neveu model is discussed.

The quantization conditions for the densities of states
$P_{a,j}=\rho_{a,j}+\tilde{\rho}_{a,j}$ are then
\begin{eqnarray}
2\pi P_{a,0}(\theta)&=&m_a\cosh\theta+\sum_{b=1}^P{\cal Y}_{ab}^{(P+1/2)}
\star \rho_{b,0}(\theta)-\sum_{j=1}^\infty \sigma_j^{(\infty)}
\star \tilde{\rho}_{a,j}(\theta) \nonumber\\
2\pi P_{P,0}&=&m\cosh\theta +\sum_{b=1}^P {\cal Y}^{(P+1/2)}_{Pb}\star
\rho_{b,0}(\theta)-\sum_{j=1}^\infty
\sigma_{j/2}^{(\infty)}\star \tilde{\rho}_{P,j}(\theta)-\xi\star
\tilde{\tau}(\theta)
\label{hubiii}
\end{eqnarray}
The last terms in equation (\ref{hubiii}) are expressions of the
logarithmic derivatives of the transfer matrix ${\cal T}$ eigenvalues.
The kernels ${\cal Y}_{ab}$ are simply related with the S matrices
described in the previous sections. Except for the case $a=b=P$ given
above in (\ref{Ypp}), one has ${\cal Y}_{ab}^{(P+1/2)}={1\over
i}{d\over d\theta} \ln S_{ab}$, where $S_{ab}$ is the prefactor for
scattering particles belonging to the representations with highest
weights $\mu_a$ and $\mu_b$.  As explained above, these $S$ matrix
elements do not probe the kink structure, so the same computation will
give these kernels for $O(N)$ for even and odd $N$. Thus these 
kernels for odd $N$ can be read off from the $N$ even results in
\cite{meTBA}, giving
\begin{eqnarray}
\hat{{\cal Y}}^{(P+1/2)}_{ab}(\omega)&=&\delta_{ab}-e^{|\omega|} \frac{\cosh
  ((P-1/2-max(a,b))\omega)~\sinh (min(a,b)\omega)}{
  \cosh((P-1/2)\omega)~\sinh(\omega)}\nonumber\\
\hat{{\cal Y}}^{(P+1/2)}_{a,P}&=&-e^{|\omega|} \frac{\sinh (a\omega)}{
  2\cosh((P-1/2)\omega)~\sinh(\omega)}\nonumber\\
\hat{\cal{Y}}^{(P+1/2)}_{P,P}&=&1-e^{|\omega|} \frac{\sinh (P\omega)}{
4\cosh((P-1/2)\omega)\sinh(\omega)\cosh(\omega/2)}\label{hubii}
\end{eqnarray}
for $a,b=1\dots P-1$. The latter kernel came from (\ref{Ypp}).
The pseudoparticle densities are given by
the auxiliary Bethe system
\begin{eqnarray}
2\pi \rho_{a,j}(\theta)&=&\sigma_j^{(\infty)}\star
\rho_{a,0}-\sum_{l=1}^\infty \sum_{b=1}^{P-1}
A_{jl}^{(\infty)}\star K_{ab} \star
\tilde{\rho}_{b,l}(\theta)-\sum_{l=1}^\infty A_{j,l/2}^{(\infty)}\star
K_{aP}\star \tilde{\rho}_{P,l}\nonumber\\
2\pi\rho_{P,j}(\theta)&=&\sigma_{j/2}^{(\infty)}\star
\rho_{P,0}-\sum_{l=1}^\infty \sum_{b=1}^{P-1}
A_{j/2,l}^{(\infty)}\star K_{Pb} \star
\tilde{\rho}_{b,l}(\theta)-\sum_{l=1}^\infty A_{j/2,l/2}^{(\infty)}\star
K_{PP}\star \tilde{\rho}_{P,l}
\nonumber\\
2\pi\left(\tau+\tilde{\tau}\right)&=&\xi\star \rho_{P,0}
\label{hubiv}
\end{eqnarray}
where the kernel $K$ in Fourier space is
\begin{eqnarray}
\hat{K}_{a,a\pm 1}&=&-{1\over  2\cosh\omega}\qquad \hat{K}_{aa}=1\nonumber\\
\hat{K}_{P,P-1}&=&{\cosh\omega/2\over\cosh\omega}\qquad \hat{K}_{PP}={\coth
  \omega/2\over\coth\omega}.
\label{hubv}
\end{eqnarray}
The equations (\ref{hubiv}) are the continuum limit of the $SO(2P+1)$
Bethe equations, with sources terms associated to all the $\rho_a$
representations.  In the preceding equations
(\ref{hubiii}--\ref{hubv}), the index $a$ takes values $1\dots P-1$;
the equations involving the index $P$ are given explicitly. The key
feature of this system is the appearance of factors of 2 in the terms
involving the latter, which in the Dynkin diagram corresponds to the
spinor representation. It is directly related to the fact that the
$P^{th}$ root is the shortest, and has length 1 while all the others
have length 2.

The TBA equations are written in terms of functions $\epsilon_{a,j}$, defined
as
$${\rho_{a,j}\over P_{a,j}} = {1\over 1+ e^{\epsilon_{a,j}/T}}\qquad\qquad
{\tau\over \tau + \tilde\tau} = {1\over 1+ e^{\epsilon_{P,-1}/T}}.$$
for $a=1\dots P$. The values $j$ runs over depends on the value of $a$:
for $a=1\dots P-1$, $j$ takes values $0,1,\dots\infty$, while for $a=P$,
$j=-1,0,1\dots\infty$. The extra function $\epsilon_{P,-1}$ arises from
the diagonalization of the BPS $S$ matrix.
This problem can now be put in a considerably simpler and rather
universal form by inverting the kernels.  The resulting TBA equations
are
\begin{eqnarray}
 \epsilon_{a,j}(\theta)=&-T\phi\star
 \left[\ln\left(1+e^{-\epsilon_{a,j+1}/T}\right)\right.&
\left.+\ln\left(1+e^{-\epsilon_{a,j-1}/T}\right)\right]
 \nonumber\\ &+T\phi\star
 \left[\ln\left(1+e^{\epsilon_{a+1,j}/T}\right)\right.&
\left.+\ln\left(1+e^{\epsilon_{a-1,j}/T}\right)
 \right] ,~~~~~ a=1,\ldots,P-2
\label{tba1}
\end{eqnarray}
together with
\begin{eqnarray}
 \epsilon_{P-1,j}(\theta)&=&-T\phi\star
 \left[\ln\left(1+e^{-\epsilon_{P-1,j+1}/T}\right)\right.\left.+\ln\left(1+e^{-\epsilon_{P-1,j-1}/T}\right)\right]\nonumber\\
 &&+T\phi\star
 \left[\ln\left(1+e^{\epsilon_{P-2,j}/T}\right)\right.\left.+\ln\left(1+e^{\epsilon_{P,2j+1}/T}\right)
+\ln\left(1+e^{\epsilon_{P,2j-1}/T}\right)\right]\nonumber\\
&&+T\psi \star
 \ln\left(1+e^{\epsilon_{P,2j}/T}\right)
\label{tba2}
\end{eqnarray}
and 
\begin{equation}
\epsilon_{P,j}(\theta)=-T\xi\star
 \left[\ln\left(1+e^{-\epsilon_{P,j+1}/T}\right)+\ln\left(1+e^{-\epsilon_{P,j-1}/T}\right)\right]
+T\xi \star
 \ln\left(1+e^{\epsilon_{P-1,j/2}/T}\right)
\label{tba3}
\end{equation}
In the last equation, the coupling to $\epsilon_{P-1,j/2}$ occurs only
when $j$ is even. The kernels are defined by their Fourier transform,
$$\hat{\phi}={1\over 2\cosh(\omega)}\qquad
\hat{\psi}={\cosh({\omega\over 2})\over\cosh(\omega)}\qquad
\hat{\xi}={1\over 2\cosh({\omega\over 2})}.$$ 
As usual, the mass
terms disappear from the equations, but
are encoded in the asymptotic boundary conditions
$$\epsilon_{a,0}\rightarrow m_a\cosh\theta\qquad
\epsilon_{P,0}\rightarrow m\cosh\theta,\qquad\theta\rightarrow\infty$$ The
TBA equations can be conveniently encoded in the diagram of figure
4.  The is the non-simply-laced generalization of the diagrams of
\cite{meTBA}. One remarkable feature is how the extra pseudoparticle
coming from the BPS structure (the node on the bottom left)
fits in perfectly with the pseudoparticles coming from the diagonalization of the spinor part of the $S$ matrix. In this sense one could
infer the existence of the BPS structure
from a careful examination of the structure of the TBA equations.
\begin{figure}
\centerline{\epsfxsize=4.0in\epsffile{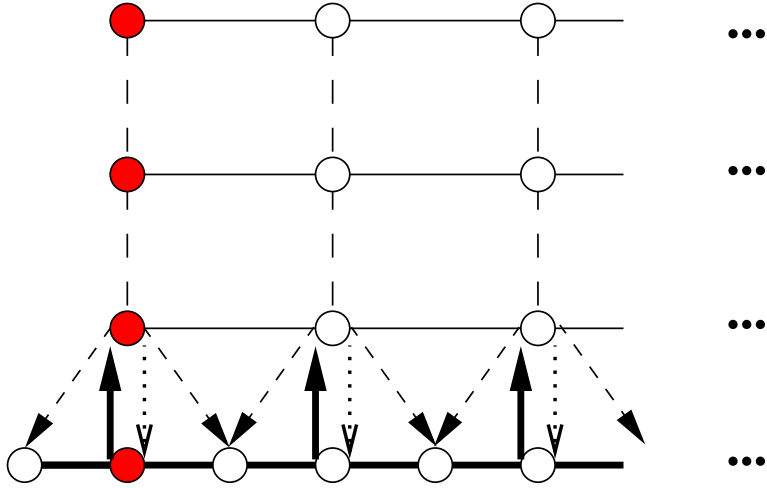}}
\bigskip\bigskip
\caption{The TBA for the $O(2P+1)$ Gross-Neveu model (here $P=4$)}
\end{figure}

The entire purpose of this section is to verify that this $S$ matrix
does give the correct free energy (\ref{freeuv}) in the ultraviolet $m\to 0$. 
This
is of course a  major  check: it shows that this  complicated
spectrum and $S$ matrix conspire to give the correct central charge $c=P+1/2$ in the UV,
corresponding to $2P+1$ Majorana fermions.
The free energy per unit length $F$
is given in terms of these dressed energies $\epsilon_a$ as
\begin{equation}
F (m,T)=
 - T \sum_a m_a \int_{-\infty}^{\infty} \frac{d\theta}{2\pi} \cosh\theta
\, \ln\left( 1 + e^{-\epsilon_{a,0}(\theta)/T}\right)
\label{freea}
\end{equation}
By rewriting $F$ as $m\to 0$ in terms of dilogarithms, we have verified that
one does indeed obtain $c=P+1/2$ as required. Thus we take it as proven
that this is the correct $S$ matrix.

Another check is that
in the IR limit $m_a\to\infty$, the correct particle multiplicities
are obtained. Namely, one should have
\begin{equation}
\lim_{m_a\to\infty} F (m,T) = 
-T \sum_a n_a m_a\int_{-\infty}^{\infty} \frac{d\theta}{2\pi} \cosh\theta\,
e^{-m_a\cosh(\theta)/T}
\label{freeir}
\end{equation}
where $n_a$ is the number of particles with mass $m_a$. 
To obtain this by taking the limit of (\ref{freea}) takes a little bit
of work. Define
\begin{eqnarray*}
Y_{a,j} &\equiv& \lim_{\theta\to\infty} e^{\epsilon_{a,j}(\theta)}
\qquad \qquad j\ge 1 \\
Y_{a,0} &\equiv& \lim_{\theta\to\infty} e^{\epsilon_{a,0}(\theta)-m_a\cosh\theta}
\end{eqnarray*}
It then follows from (\ref{freea},\ref{freeir}) that $n_a = Y_{a,0}$. To find
the $Y_{a,j}$ requires taking the $m\to\infty$ limit of the TBA equations
(\ref{tba1},\ref{tba2},\ref{tba3}). In this limit, one can replace the $\epsilon_{a,j}(\theta)$ by $\ln
Y_{a,j}$. Thus the integrals can be done explicitly,
giving a set of polynomial equations for the $Y_{a,j}$. For example,
for $j=0$ they are
\begin{eqnarray*}
(Y_{a,0})^2 &=& Y_{a-1,0}Y_{a+1,0}(1+Y_{a,1})\qquad\qquad a=1,\dots ,P-2\\
(Y_{P-1,0})^2 &=& Y_{P-2,0}(Y_{P,0})^2 (1+Y_{P-1,1}) Y_{P,1}(1 + Y_{P,1})^{-1}
Y_{P,-1}(1 + Y_{P,-1})^{-1}
\\
(Y_{P,0})^2 &=& Y_{P-1,0}(1+Y_{P,-1})(1+Y_{P,1})
\end{eqnarray*}
where $Y_{0,0}\equiv 1$.
The (not written) equations for $Y_{a,j}$ for $j\ge 1$ do not depend
on the $Y_{a,0}$, so these can be solved separately. One finds
immediately that $Y_{P,-1}=1$, but unfortunately, we were not able to
derive an explicit closed-form solution for the other $Y_{a,j}$. 
However, it is easy to find them by solving the polynomial
equations numerically. For example, for $P=2$, one finds that
$Y_{1,1}=14/11$, and $Y_{2,1}=11/5$. Plugging this into the equation
for $Y_{a,0}$ gives $Y_{1,0}=n_1=5$ and $Y_{2,0}=n_2=4\sqrt{2}$.
These indeed are the correct multiplicities for $P=2$.
For $P=3$, we find $Y_{1,1}=27/22$, $Y_{2,1}=95/147$ and
$Y_{3,1}=21/11$. This yields $n_1=7$, $n_2=22$ and $n_3=8\sqrt{2}$.
In particular, note that $n_{2}=21+1$, the dimensions of the antisymmetric
and singlet representations of $O(7)$.
This checks that the transfer matrix
diagonalization in \cite{Kuniba} indeed includes all the particles at
each mass, even though they come from more than one representation for
$a=2\dots P-1$.

As another check, we can easily generalize this TBA calculation
to the $SO(2P+1)_k\times SO(2P+1)_1/SO(2P+1)_{k+1}$ coset models mentioned
at the end of the last section.
The TBA is almost the same as that above, except that the
right hand side is truncated, so the $j$ in $\epsilon_{a,j}$ 
runs only from $0\dots k-1$ for $a=1,\ldots,P-1$, while
$j=-1,0,1\dots 2k-1$ for the $a=P$ nodes \cite{Kuniba}.
One can check that for this truncated TBA system 
$$c_{UV} = \frac{k(2P+1)(4P-1+k)}{2(k+2P-1)(k+2P)}$$
as required.
For $k=1$, the conformal theory has central charge
$c=1$, and the coset perturbation coincides with the sine-Gordon model
for ${\beta_{SG}^2\over 8\pi}={2\over 2P+1}$. This TBA is
represented by the diagram in figure 5, and was first studied in
\cite{Itoyama,Tateo}.
\begin{figure}
\centerline{\epsfxsize=1.0in\epsffile{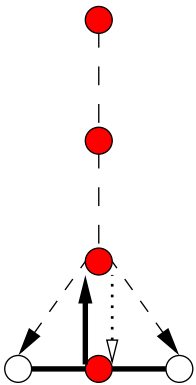}}
\bigskip\bigskip
\caption{The TBA for the sine-Gordon model
 at ${\beta_{SG}^2\over 8\pi}={2\over 2P+1}$ (here $P=4$)}
\end{figure}

\section{Conclusion}

We have completed the solution of the Gross-Neveu model for any number
$N$ of fermions. For odd $N$, generalized supersymmetry results in the
existence of BPS kinks. We found the exact $S$ matrix for these kinks,
and used this to compute the exact free energy.

One striking feature is that there are a non-integer number $K=2^{N/2}$ of
these kinks, in the sense that the number of $n$-kink states goes as
$K^n$.  In the simplest case, the generalized supersymmetry reduces to
${\cal N}=1$ supersymmetry in two dimensions. We note that a number of
results concerning BPS kinks in 1+1 dimensional supersymmetric field
theories have been derived recently (see \cite{Losev} and references
therein), but to our knowledge, none of these papers have discussed
this multiplet with the $\sqrt{2}$ particles. However, 
 unusual particle statistics related to Clifford algebras have
been discussed in \cite{Finkelstein}.

We have generalized these results to a large number of models with
four-fermion interactions \cite{us}. These models include
Gross-Neveu-like models with $Sp(2N)$ symmetry, and multi-flavor
generalizations of the Gross-Neveu model. As in the $O(N)$ case, kinks
with non-integer $K$ appear whenever the symmetry algebra is not
simply laced.

\bigskip\bigskip 

The work of P.F. is supported by a DOE OJI Award, a Sloan
Foundation Fellowship, and by NSF grant DMR-9802813. The work of
H.S. is supported by the DOE and the NSF through the NYI program.

%
\renewcommand{\baselinestretch}{1}

\end{document}